\documentclass{aa}  

\usepackage{graphicx}
\usepackage{txfonts}
\usepackage{subfig}
\begin{document}

   \title{Detection of faint stars near Sgr~A* with GRAVITY}


   \author{GRAVITY Collaboration\thanks{GRAVITY is developed
    in a collaboration by the Max Planck Institute for
    extraterrestrial Physics, LESIA of Observatoire de Paris/Universit\'e PSL/CNRS/Sorbonne Universit\'e/Universit\'e de Paris and IPAG of Universit\'e Grenoble Alpes /
    CNRS, the Max Planck Institute for Astronomy, the University of
    Cologne, the CENTRA - Centro de Astrofisica e Gravita\c c\~ao, and
    the European Southern Observatory. $\,\,\,\,\,\,$ Corresponding author: F. Gao 
    (email: fgao$@$mpe.mpg.de) and T. Paumard (email: thibaut.paumard$@$obspm.fr)
    }:
R.~Abuter\inst{8}
\and A.~Amorim\inst{6,12}
\and M.~Baub\"ock\inst{1}
\and J.P.~Berger\inst{5,8}
\and H.~Bonnet\inst{8}
\and W.~Brandner\inst{3}
\and Y.~Cl\'{e}net\inst{2}
\and Y.~Dallilar\inst{1}
\and R.~Davies\inst{1}
\and P.T.~de~Zeeuw\inst{10,1}
\and J.~Dexter\inst{13,1}
\and A.~Drescher\inst{1,16}
\and F.~Eisenhauer\inst{1}
\and N.M.~F\"orster~Schreiber\inst{1} 
\and P.~Garcia\inst{7,12}
\and F.~Gao\inst{1}
\and E.~Gendron\inst{2}
\and R.~Genzel\inst{1,11}
\and S.~Gillessen\inst{1}
\and M.~Habibi\inst{1}
\and X.~Haubois\inst{9}
\and G.~Hei{\ss}el\inst{2}
\and T.~Henning\inst{3}
\and S.~Hippler\inst{3}
\and M.~Horrobin\inst{4}
\and A.~Jim\'enez-Rosales\inst{1}
\and L.~Jochum\inst{9}
\and L.~Jocou\inst{5}
\and A.~Kaufer\inst{9}
\and P.~Kervella\inst{2}
\and S.~Lacour\inst{2}
\and V.~Lapeyr\`ere\inst{2}
\and J.-B.~Le~Bouquin\inst{5}
\and P.~L\'ena\inst{2}
\and D.~Lutz\inst{1}
\and M.~Nowak\inst{15,2}
\and T.~Ott\inst{1}
\and T.~Paumard\inst{2}
\and K.~Perraut\inst{5}
\and G.~Perrin\inst{2}
\and O.~Pfuhl\inst{8,1}
\and S.~Rabien\inst{1}
\and G.~Rodr\'iguez-Coira\inst{2}
\and J.~Shangguan\inst{1}
\and T.~Shimizu\inst{1}
\and S.~Scheithauer\inst{3}
\and J.~Stadler\inst{1}
\and O.~Straub\inst{1}
\and C.~Straubmeier\inst{4}
\and E.~Sturm\inst{1}
\and L.J.~Tacconi\inst{1}
\and F.~Vincent\inst{2}
\and S.~von~Fellenberg\inst{1}
\and I.~Waisberg\inst{14,1}
\and F.~Widmann\inst{1}
\and E.~Wieprecht\inst{1}
\and E.~Wiezorrek\inst{1} 
\and J.~Woillez\inst{8}
\and S.~Yazici\inst{1,4}
\and G.~Zins\inst{9}
}

\institute{
Max Planck Institute for extraterrestrial Physics,
Giessenbachstra{\ss}e~1, 85748 Garching, Germany
\and LESIA, Observatoire de Paris, Universit\'e PSL, CNRS, Sorbonne Universit\'e, Universit\'e de Paris, 5 place Jules Janssen, 92195 Meudon, France
\and Max Planck Institute for Astronomy, K\"onigstuhl 17, 
69117 Heidelberg, Germany
\and $1^{\rm st}$ Institute of Physics, University of Cologne,
Z\"ulpicher Stra{\ss}e 77, 50937 Cologne, Germany
\and Univ. Grenoble Alpes, CNRS, IPAG, 38000 Grenoble, France
\and Universidade de Lisboa - Faculdade de Ci\^encias, Campo Grande,
1749-016 Lisboa, Portugal 
\and Faculdade de Engenharia, Universidade do Porto, rua Dr. Roberto
Frias, 4200-465 Porto, Portugal 
\and European Southern Observatory, Karl-Schwarzschild-Stra{\ss}e 2, 85748
Garching, Germany
\and European Southern Observatory, Casilla 19001, Santiago 19, Chile
\and Sterrewacht Leiden, Leiden University, Postbus 9513, 2300 RA
Leiden, The Netherlands
\and Departments of Physics and Astronomy, Le Conte Hall, University
of California, Berkeley, CA 94720, USA
\and CENTRA - Centro de Astrof\'{\i}sica e
Gravita\c c\~ao, IST, Universidade de Lisboa, 1049-001 Lisboa,
Portugal
\and Department of Astrophysical \& Planetary Sciences, JILA, Duane Physics Bldg., 2000 Colorado Ave, University of Colorado, Boulder, CO 80309, USA
\and Department of Particle Physics \& Astrophysics, Weizmann Institute of Science, Rehovot 76100, Israel
\and Institute of Astronomy, Madingley Road, Cambridge CB3 0HA, UK
\and Department of Physics, Technical University Munich, James-Franck-Straße 1,  85748 Garching, Germany
}

   \date{Received 28 September 2020; accepted 4 November 2020}

 
  \abstract{ 
  The spin of the supermassive black hole that resides at the Galactic Centre can in principle be measured by accurate measurements of the orbits of stars that are much closer to Sgr~A* than S2, the orbit of which recently provided the measurement of the gravitational redshift and the Schwarzschild precession. The GRAVITY near-infrared interferometric instrument combining the four 8m telescopes of the VLT provides a spatial resolution of 2-4 mas, breaking the confusion barrier for adaptive-optics-assisted imaging with a single 8-10m telescope. We used GRAVITY to observe Sgr~A* over a period of six months in 2019 and employed interferometric reconstruction methods developed in radio astronomy to search for faint objects near Sgr~A*. This revealed a slowly moving star of magnitude 18.9 in K band within 30mas of Sgr~A*. The position and proper motion of the star are consistent with the previously known star S62, which is at a substantially larger physical distance, but in projection passes close to Sgr~A*. Observations in August and September 2019 easily detected S29, with K-magnitude of 16.6, at approximately 130 mas from Sgr~A*. The planned upgrades of GRAVITY, and further improvements in the calibration, hold the promise of finding stars fainter than magnitude 19 at K.
}


   \maketitle
%

\section{Introduction}

   The Galactic Center (GC) excels as a laboratory for astrophysics and general relativity (GR) around a massive black hole (MBH, \citealt{2010RvMP...82.3121G}). The observation of stellar orbits in the GC around the radio source Sgr~A* \citep{1996Natur.383..415E, 1998ApJ...509..678G, 2002Natur.419..694S, 2003ApJ...586L.127G, 2008ApJ...689.1044G, 2009ApJ...692.1075G, 2017ApJ...837...30G} has opened a route to testing gravity in the vicinity of a MBH with clean test particles. The gravitational redshift from Sgr~A* has been seen at high significance in the spectrum of the star S2 during the 2018 pericenter passage on its 16-year orbit \citep{2018A&A...615L..15G, 2019A&A...625L..10G, 2019Sci...365..664D}. Recently the relativistic Schwarzschild precession of the pericenter has been detected in S2's orbit as well \citep{2020A&A...636L...5G}. The effects detected are of order $\beta^2$ where $\beta = v/c$ \citep{2006ApJ...639L..21Z}. It is not clear, however, whether higher order effects such as the Lense-Thirring precession due to the spin of the black hole can be detected in the orbit of S2, since these fall off faster with distance from the MBH. In addition, stars located farther away from Sgr~A* will be more affected by Newtonian perturbations by surrounding stars and dark objects, which can make a spin measurement with S2 more difficult \citep{2010PhRvD..81f2002M, 2017ApJ...834..198Z}. Hence, it is natural to look for stars at smaller radii. In particular, such a star could offer the possibility of measuring Sgr~A*'s spin \citep{2018MNRAS.476.3600W}. If it were possible to additionally measure the quadrupole moment of the MBH in an independent way, the relation between these two parameters would constitute a test of the no-hair theorem of GR \citep{2008ApJ...674L...25W, 2018MNRAS.476.3600W}. 
   
  Beyond the purpose of testing GR, the GC cluster is the most important template for galactic nuclei. These are the sites of extreme mass ratio inspirals \citep{2007CQGra..24R.113A}, one of the source categories for the gravitational wave observatory LISA \citep{2019arXiv190706482B}. Understanding the GC cluster down to the smallest possible scales will deliver important anchor points for understanding the structure and dynamics of these stellar systems, and thus for predictions of the expected event rates.

  The number of stars expected at smaller radii has until now been estimated by extrapolating the density profile in the GC to radii smaller than the resolution limit of $\sim$ $60\,$mas in the near-infrared provided by 8-10m class telescopes and at the same time extrapolating the mass function to stars fainter than the confusion limit in the central arcsecond around $m_K\approx 18$. Both functions have been determined in the literature \citep{2003ApJ...594..812G, 2013ApJ...764..154D, 2018A&A...609A..26G}, and the resulting estimate for the number of stars suitable for GR tests is of order 1 \citep{2018MNRAS.476.3600W}.

  The near-infrared interferometric instrument GRAVITY \citep{2017A&A...602A..94G} coupled to the four 8m telescopes of the VLT makes it possible to detect and trace such faint stars for the first time with an angular resolution that exceeds that of adaptive-optics assisted imaging on 8-10m telescopes by a factor of $\sim$ 20. Here, we report on the detection of a faint star ($m_K \approx 18.9 $) within $30\,$mas of Sgr~A*. We use the classical radio interferometry data reduction/image reconstruction package Astronomy Image Process Software (AIPS, \citealt{2003ASSL..285..109G}) for our work. This detection does not yet explore the signal-to-noise ratio limit of GRAVITY, but is limited by our ability to model the point spread function (PSF) of the sparse 4-telescope interferometer and the presence of other sources and their sidelobes in the field-of-view (FOV).
  

  This paper is organized as follows: We describe our observation in Section 2. Then in Section 3 we describe in detail the data reduction and image reconstruction process to derive the images with Sgr~A* removed. We show the faint star detection in Section 4, together with validation from model fitting and constraints on the proper motion of the detected star. We cross-check our detection with the expected positions of known S-stars and also give the limitation of our current imaging technique in Section 5. Finally, we give our conclusion in Section 6.

   \begin{figure}
   \centering
   \includegraphics[width=.9\hsize]{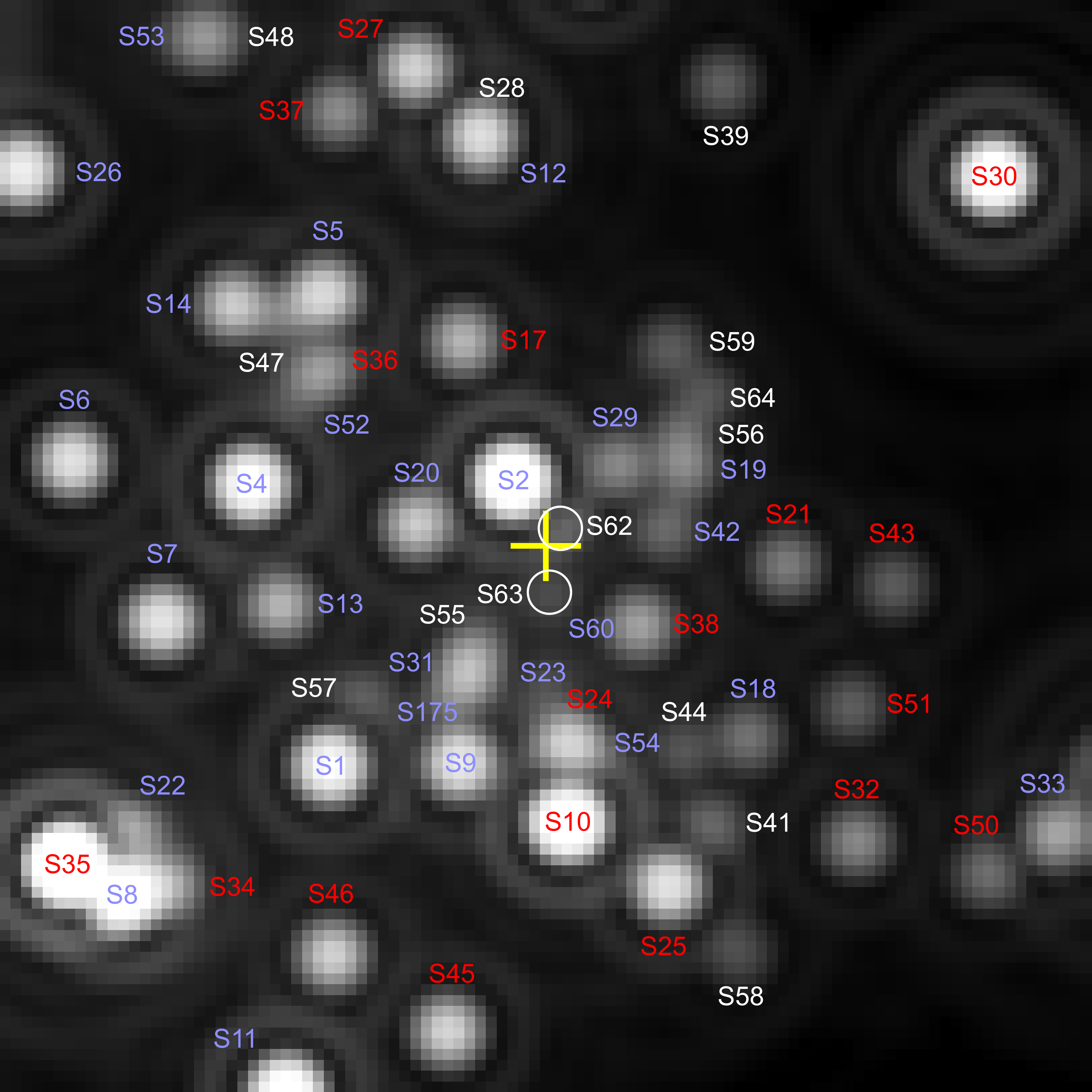}
      \caption{Simulated image of the central 1.3'' for the epoch 2019.5. with the latest motions laws of the S-stars as obtained from VLT/NACO observations until end 2019, this image shows the expected appearance of the vicinity of Sgr~A* at the resolution and pixel scale of the adaptive optics imager NACO at the VLT. Red labels indicate late-type stars and blue labels indicate early-type stars. White labels indicate no spectral identification is available.
              }
         \label{figure:naco-mosaic}
   \end{figure}

\section{Observations}

   We observed Sgr~A* and the immediate surrounding (Fig.~\ref{figure:naco-mosaic}) using the VLT/GRAVITY instrument at ESO Paranal 
   Observatory between 2019 March 27th and 2019 September 15th, under the GTO program ID 0102.B-0689 and 0103.B-0032, spread over a total of 28 nights. The data are the same as those used in \citet{2020A&A...636L...5G}. Compared to the 2017 and 2018 data on the Galactic Center, the 2019 data have the advantage that for the first time the bright star S2 has moved sufficiently far away from Sgr~A* that its sidelobes do not dominate the residual structure in CLEANed \footnote{Throughout this paper, we use the term "CLEANed" image to denote a deconvolved image after applying the CLEAN algorithm.} images. 
   
   We used the low spectral resolution and the split polarization mode. The nearby star IRS~7 was used for the adaptive optics correction, and we chose IRS~16C within
   the 2" VLTI field of view for fringe-tracking with the highest tracking rate of 1kHz. We list the observing dates and numbers of Sgr~A* frames used for imaging in Table~\ref{t1}.   
   
    \begin{table}
    \caption{\label{t1}Observation details}
    \centering
    \begin{tabular}{lccc}
    \hline\hline
    Date & $N_{obs}$  &$N_{used}$  &Incl.  \\
    \hline
    2019-03-27      & 6      &  0    & N \\
    2019-03-28      & 8      &  0    & N \\
    2019-03-30      & 2      &  0    & N \\
    2019-03-31      & 5      &  0    & N \\
    \hline
    2019-04-15      & 8      &  0    & N \\
    2019-04-16      & 4      &  0    & N \\
    2019-04-18      & 17     &  17    & Y \\
    2019-04-19      & 12     &  10    & Y \\
    2019-04-21      & 28     &  11    & N \\
    \hline
    2019-06-13      & 4      &  0    & N \\
    2019-06-14      & 4      &  0    & N \\
    2019-06-16      & 17     &  4    & N \\
    2019-06-17      & 5      &  0    & N \\  
    2019-06-18      & 4      &  0    & N \\
    2019-06-19      & 8      &  6    & N \\
    2019-06-20      & 27     &  15   & Y \\ 
    \hline
    2019-07-15      & 8      & 7     & Y \\
    2019-07-17      & 39     & 32    & Y \\
    \hline
    2019-08-13      & 21      &  21    & Y \\
    2019-08-14      & 10      &  8     & Y \\
    2019-08-15      & 25      &  25    & Y \\
    2019-08-17      & 8       &  0     & N \\  
    2019-08-18      & 18      &  14    & Y \\
    2019-08-19      & 21      &  15    & Y \\
    \hline
    2019-09-11      & 9       &  8     & Y \\
    2019-09-12      & 14      &  8    & Y \\
    2019-09-13      & 11      &  5     & Y \\  
    2019-09-15      & 12      &  10    & Y \\
    \hline
    \end{tabular}
    \tablefoot{Observation details. Here we list all the Sgr~A* observation taken in 2019, together with number of frames observed ($N_{obs}$), number of frames used in image per night ($N_{used}$) and whether they are included in the final image per month. Each frame is 320 seconds long.}
    \end{table}

   During most of our observations, we centered the science fiber on Sgr~A*.
   The science fiber has an acceptance angle of $\approx 74\,$mas full width half maximum. Each frame lasted 320 seconds, composed of 32 exposures of 10 seconds. During each night's observation we also pointed the science fiber to the
   nearby bright stars S2 and R2 ($m_{K}\approx12.1$ mag, separation $\approx$ 1.5$"$), bracketing the Sgr~A* observing blocks for later
   calibration purposes. 

   On the nights of August 13th, 14th and September 13th 2019, in addition to the
   standard Sgr~A* observing sequence, we also pointed the science fiber at $(\mathrm{RA}, \mathrm{Dec}) \approx (-86, +90)\,$mas relative to Sgr~A* to check the potential detection of a proposed short period faint star as reported
   by \cite{Peissker}.

\section{Data reduction, calibration and image reconstruction}

   The data reduction consists of several steps. We reduce each
   night's data separately. The first step is to use the standard GRAVITY pipeline \citep{2014SPIE.9146E..2DL}
   to calibrate the instrumental response and derive calibrated 
   interferometric quantities from the raw data.
   We follow the default pipeline settings except for the calibration
   step, where we only use a single, carefully chosen S2 frame to calibrate the Sgr~A*
   data for each night.

   \subsection{Data format conversion}
   
   The standard GRAVITY data products after calibration are stored in the OIFITS format \citep{2017A&A...597A...8D}. In order to benefit from existing radio interferometry imaging reconstruction software and algorithms, we here converted the GRAVITY data product into the UVFITS format (\citealt{2003ASSL..285..109G}), which is commonly used in radio interferometry. 
   
   The conversion was done with a python script which we adapted from the EHT-imaging package \citep{2019ApJ...875L...4E}. The essential data we read in from the calibrated OIFITS files are the VISAMP, VISPHI, MJD and U-V coordinates (UCOORD, VCOORD) columns from the OI$\_$VIS table for the science channel output together with the corresponding OI$\_$FLUX table. We then rewrite these following the UVFITS format convention. We also take the OI$\_$WAVELENGTH table and write it into the FQ table used in UVFITS. Since the effective bandwidth for each wavelength output is different, we put each wavelength output as an independent intermediate frequency (IF) rather than an independent channel in the UVFITS file.  
   
   \subsection{Additional amplitude calibration}

   Before writing out the UVFITS data product, we re-calibrate and re-scale the visibility amplitudes with
   the photometric flux of each baseline calculated from each telescope pair (using the OI$\_$FLUX table). This allows us to get the correlated flux from each baseline instead of the default normalized visibility, which does not reflect the true brightness of the target. During this step, we also correct for the attenuation of the telescope flux due to different air mass and AO correction by fitting a polynomial function to the OI$\_$FLUX of several S2 frames per telescope across one night and interpolating the correction per telescope accordingly.
   
   We then renormalize each visibility amplitude with that of the S2 frame used for calibration in the previous step. Thus, our visibility amplitudes are normalized such that a visibility amplitude of 1 equals
   S2's magnitude in the Ks band (m$_{K}$ = 14.1, \citealt{2017ApJ...837...30G}).
   
   The final VIS$\_$AMP quantity we write out is: 
   \begin{equation}
    a_{ij,final}^{obj(t)} = a_{ij}^{obj(t)} \sqrt{ \frac{ {f_i(t)f_j(t)\sigma_i(t)\sigma_j(t)}} { f_i(t_0)f_j(t_0)\sigma_i(t_0)\sigma_j(t_0)}},  
   \end{equation}
   
   \noindent where \(a_{ij}^{obj(t)}\) is the pipeline-produced visibility amplitude of a certain target frame at time $t$  between telescope $i$ and $j$, \(f_i(t)\) is the measured flux from telescope $i$ at time $t$, and \(\sigma_i(t)\) is the fitted air mass correction coefficient for telescope $i$  at time $t$ . The time \(t_0\) corresponds to the S2 frame used for calibration in the GRAVITY pipeline. 
   
   \begin{figure*}
   \centering
   \includegraphics[width=\hsize]{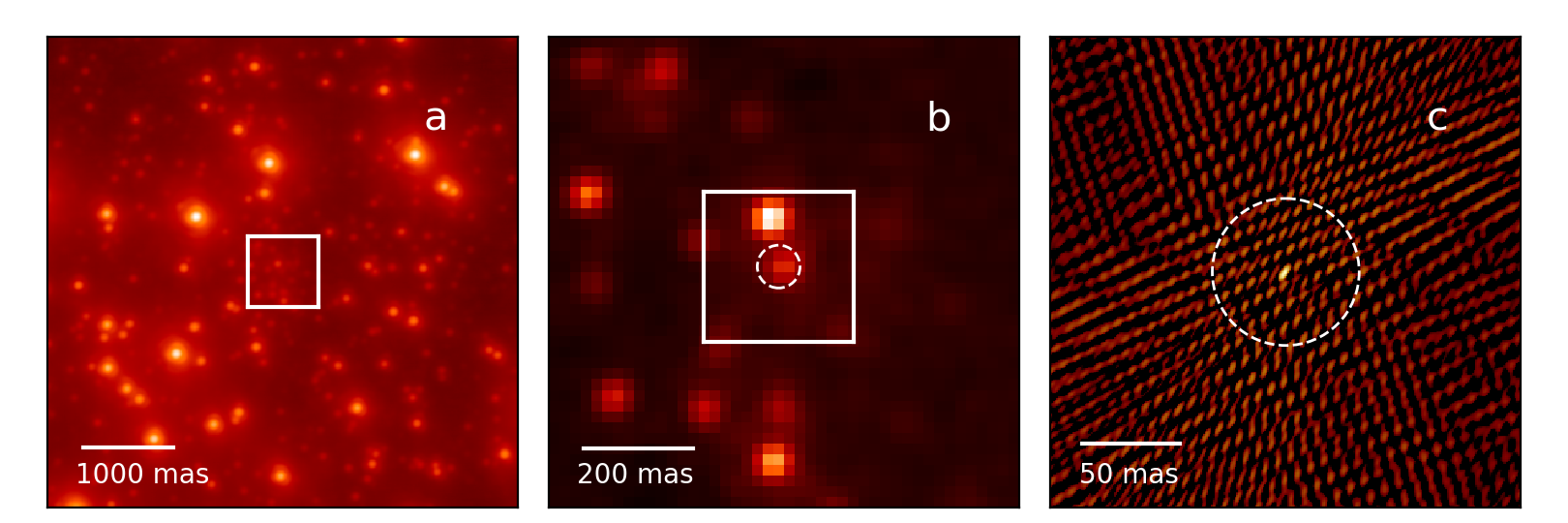}
      \caption{(a) The Galactic Center region of 5 $\times$ 5 arcsec as seen by NACO at 2.2 micron. The central white box indicates the region shown in panel b. (b) Image of the Galactic Center region of 0.8 $\times$ 0.8 arcsec from the GRAVITY acquisition camera in H-band. The central white box indicates the region shown in panel c. (c) Image of the central 240 $\times$240 mas region with Sgr~A* deconvolved with the CLEAN algorithm and overlaid on the residual background. The dashed circle in both panel b and c indicates the GRAVITY fiber FOV with an HWHM of $\sim$74 mas. S2 was outside of this region and it's flux was reduced by fiber damping by about a factor of 100. 
              }
         \label{figure:example-beam-resi}
   \end{figure*}

   \subsection{Image reconstruction and deconvolution with CLEAN}
   
   After the data reduction steps, we load the calibrated 
   and amplitude renormalized complex visibility data into AIPS for image reconstruction with the task IMAGR.
   Sgr~A* is known to exhibit
   variability on a time scale of minutes in the near-infrared \citep{2003Natur.425..934G, 2004ApJ...601L.159G, 2006ApJ...640L.163G, 2008A&A...492..337E, 2009ApJ...691.1021D, 2011ApJ...728...37D, 2018ApJ...863...15W, 2020A&A...636L...5G, 2020arXiv200407185T}.
   This variability must be taken into account during image reconstruction in order to remove as much as possible
   the sidelobes of Sgr~A*, which dominate the residual map and limit the sensitivity. We reduce the impact of the variability by reconstructing each 320 seconds long frame individually. While the variability on this time scale can be high during flares, it is much lower during quiescent phases. Consequently, we assume the flux to be constant over the duration of each frame.
   
   In the spectral domain, we only include data from $2.111\,\mu$m to $2.444\,\mu$m in order to minimize systematic biases from the edge channels, which are subject to increased instrumental (short wavelengths) and thermal (long wavelengths) background. We use the data from both polarizations (as Stokes I) in the reconstruction of the images. Given that our synthesized beam
   size is about $2\times4\,$mas, we choose a pixel size of $0.8\,$mas and an image size of $512\times 512$ pixels, which can fully cover the fiber FOV with a FWHM of about 74 mas. When applying CLEAN to the Sgr~A* frames, we applied a spectral index of +6. 
   \footnote{We only keep frames where Sgr~A* is relatively faint for later analysis, in which case Sgr~A* is expected to have a near-infrared spectral index ($\nu L_{\nu} \sim \nu^\alpha$) between -3 and -1 \citep{2010RvMP...82.3121G, 2018ApJ...863...15W}. Since we use S2 data to normalize the Sgr~A* data, we take into account a spectral index of +2 from the Rayleigh-Jeans approximation of the black body radiation from S2: this would give us a spectral index range of $\alpha$-1 between -6 and -4. We then choose the value of -6 as to cover the extreme cases and finally we flip the sign to follow the spectral index definition in AIPS to get the +6 value.}
   The IMAGR task requires the user to choose the data weighting via a parameter called ROBUST, which we set to zero. This ensures a balanced result between natural and uniform weighting. The outputs at this stage are so-called "dirty images", which in theory are the convolution of the source distribution with the beam pattern.
   
   We found the following procedure effective for creating residual images, in which we can identify additional sources beyond Sgr~A*, from the dirty images: \\
   $\bullet$ CLEAN. Per 320 seconds frame, we used a small pre-defined clean box centered on Sgr~A*---the central and brightest spot---with a size of $5\times6$ pixels (i.e. smaller than the spatial periodicity of the beam pattern and thus avoiding any sidelobes). We iterate until the first negative CLEAN component appears. We repeat the same process for all the frames in a certain night to get a series of CLEANed images and residual images.\\
   $\bullet$ Selection. We then visually check the CLEANed image and the residual image with the following two criteria: 1) the CLEANed image shows no signs of abnormal instrumental behavior or strong baseline pattern for a specific baseline. 2) the peak value on the residual image is below 0.8 mJy. Due to the limited dynamic range we can reach with the CLEAN algorithm here, the 2nd criterion limits us to only those frames where Sgr~A* is relatively faint (below $\sim$2 mJy).\\
   $\bullet$ Subtraction of Sgr~A*. After that, we subtract the respective CLEAN component model from each frame in the visibility domain with the AIPS task UVSUB.\\
   $\bullet$ Combine and re-image Sgr~A*-removed data for one night. Further, we group the data by each night and combine them in the uv domain with the AIPS task DBCON. Then we re-image them to get Sgr~A*-removed images per night. Before we combine the data per month, we first cross-check these images with the criteria mentioned in the next section. 
   

   We note that we also tried to clean the images with the phase-only self-calibration by using the task CALIB. However, this did not improve our residual image in general, so no self-calibration was included in our data reduction process.
   
   As an example, we show in Fig.~\ref{figure:example-beam-resi} the GC region as observed by the VLT NACO instrument, the GRAVITY acquisition camera and one Sgr~A* reconstructed image.

   \subsection{Sanity checks}
    Apart from imaging Sgr~A* frames, we also repeat the same imaging procedure on S2 frames and R2 frames for each night (with a spectral index of 0) as a sanity check for the instrument behavior. This allows us to check our flux calibration as the magnitude of both S2 and R2 are known and they are not variable. Additionally, we can use these data to gauge the dynamic range we could reach with our method and compare that to the Sgr~A* frames.

\section{Results}

    \subsection{Source identification}
    
    We first estimate in section 4.1.1 how much on-sky movement we might expect from a star close to Sgr~A*, which puts a limit on how much data we can combine in time before our images are smeared. Then we describe the image co-adding in section 4.1.2.

    \subsubsection{Prerequisite requirement on the movement of a tentative star}
    
    A star belonging to the Sgr~A* system is gravitationally bound to the massive black hole, and hence it is useful to know the escape velocity for stars within our field of view. For a star located $10\,$mas away from Sgr~A*, the escape velocity is 
    $\approx 10^{4}\,$km~s$^{-1}$, corresponding to $\approx 250\,\mathrm{mas~yr^{-1}}$. The day-to-day motion of the star on the sky is smaller than $0.7\,$mas, and smaller than $20\,$mas from month to month. Hence, a 
    tentative star should show up in the images of multiple nights in a month in the same position, given that our synthesis beam size is about 2 $\times$ 4 mas. From one month to the next, the star might move by at most a few times the beam size. We can thus co-add the frames per month.
    
    \subsubsection{Co-adding}
    
    
    
\begin{figure*}
    \centering
    \subfloat[]{\includegraphics[width=.75\hsize]{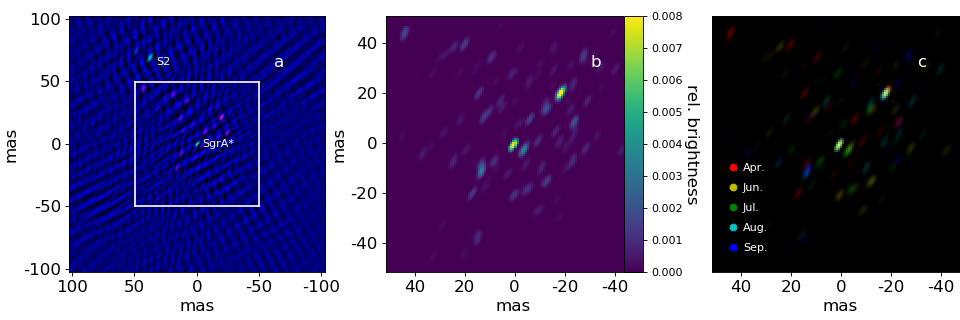}}
    
    \centering
    \subfloat[]{\includegraphics[width=\hsize]{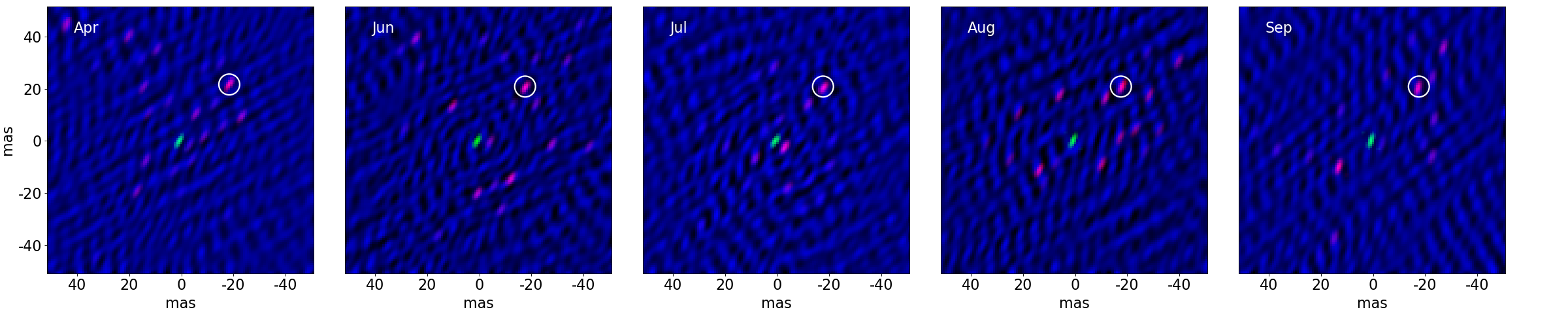}}
    \caption{Top row: (a) Three-color composite image of the full 200 mas region for the image reconstruction from the April 2019 data, shown as a typical example. Here Sgr~A* (located at the image center) and S2 are color-coded in green, other potential targets are shown in red and the residual map is shown in blue. The central white box indicates the central 100 mas region shown in other panels. (b) Stacking of the central 100 mas region around Sgr~A* for April, June, July, August and September in 2019. Top left: Intensity map of the averaged image over 5 months. In both images, the central bright spot at [0,0] mas is Sgr~A*, with its flux reduced by 40 times. (c) Image from each month is shown in a different color, illustrating the motion in the south-east direction. Bottom row: three-color composite images of the central 100 mas region around Sgr~A* for April, June, July, August and September in 2019. In each map Sgr~A* is color-coded in green, other potential targets are shown in red and the residual map is shown in blue. North is up and east to the left. S2 is outside of this region. We highlight the newly-detected object with a white circle.
              }
    \label{figure:image-per-month}
    \end{figure*}

  
    At this stage, we have a series of Sgr~A*-removed images per night for March, April, June, July, August and September in 2019. We first cross-check the multiple images for each month, knowing that a real source would not move between the nights within a given month. We deselect frames which visually deviate strongly from the monthly sample. This is a very helpful selection criterion, since the total number of Sgr~A* frames per night varies greatly, and we do not always reach the same sensitivity level. Since we reset our instrument every night, instrumental misbehavior or bad observing conditions are more likely to affect the image from a single night only. Finally, we average the remaining images to derive the best Sgr~A* removed image per month, shown in Fig.~\ref{figure:image-per-month}.
    
    We have tried to run the CLEAN algorithm on these Sgr~A*-removed images, on both the manually and automatically selected peak positions. However, the CLEAN algorithm does not finish successfully in several instances, and hence we cannot robustly CLEAN these images. We speculate that this is because the dirty image pattern on the Sgr~A*-removed images sometimes shows distortion due to time-variable systematic errors, which prevent the CLEAN algorithm from converging. Hence, we keep these Sgr~A*-removed, but not further deconvolved images for a manual analysis in the next step.
    
    \subsection{Inspection of monthly co-adds}
    The monthly co-added images show multiple patches with some bright features embedded. These are a combination of any targets in the field convolved with our interferometric beam plus the interferometric residual background.  
    
   The most prominent bright feature on these images is the one to the northeast of Sgr~A*, which is the star S2 (with the distorting effect from bandwidth smearing, and a brightness damped by our fiber profile). Despite its distance exceeding the nominal field of view of GRAVITY, S2 is very robustly detected in all images.

   Besides S2, we noticed a bright feature to the northwest of Sgr~A* consistently showing up in all our images. This second-most prominent feature over the sample of images corresponds to another star, as we show below. We perform a 2-D Gaussian fit on this bright feature to get the position and brightness relative to Sgr~A* (see Table~\ref{t2}).  
    
    Since we use S2 frames to renormalize the Sgr~A* frames, the target brightness is also normalized to S2. The relative brightness of this tentative star is $0.0084\pm 0.0009$, which is $119\pm 16$ times or $5.2 \pm 0.1$ magnitudes fainter than S2, a 14.1 magnitude source in K-band \cite{2009ApJ...692.1075G}. The detected object is about 27 mas away from Sgr~A*, which corresponds to a fiber damping loss (described by a Gaussian function with a full width half maximum of 74 mas) of $\approx 30\%$. Taking this effect into account, the intrinsic brightness of this object is $m_\mathrm{K} = 18.9 \pm 0.1$.

    \begin{table*}
    \caption{\label{t2}Fitted position and relative brightness of the faint star}
    \centering
    \begin{tabular}{lcccccccc}
    \hline\hline
    Date of the&$\delta$ RA &$\delta$ RA err. &$\delta$ DEC. &$\delta$ DEC. err.& relative & relative & 1-$\sigma$ & number of\\
    averaged image& (mas) & (mas) & (mas) & (mas) & brightness & brightness of S2 & noise level & frames used\\
    \hline
    2019.30 (Apr.18)    & -19.18     & 0.18  & 20.78     & 0.27  & 0.009 & 0.013 & 0.0019 & 27\\
    2019.47 (Jun.19)      & -18.78     & 0.14  & 19.80     & 0.23  & 0.009 & 0.011 & 0.0018 & 15\\
    2019.54 (Jul.17)      & -18.62     & 0.26  & 19.73     & 0.30  & 0.007 & 0.009 & 0.0017 & 39\\
    2019.62 (Aug.16)      & -18.49     & 0.11  & 19.88     & 0.29  & 0.008 & 0.009 & 0.0014 & 83\\
    2019.70 (Sep.12)      & -18.14     & 0.15  & 19.57     & 0.34  & 0.009 & 0.011 & 0.0017 & 31\\
    \hline
    \end{tabular}
    \tablefoot{Fitted position of our detected target relative to Sgr~A* and the image noise level for the best Sgr~A* removed image per month. Positive sign means to the East and North of Sgr~A* in Right Ascension and Declination direction, respectively. The brightness level is a normalized quantity, with the maximum of unity, which corresponds to the brightness of S2, if it was in the field center. Here the brightness levels do not reflect the physical brightness of the stars as they are also affected by bandwidth smearing and fiber damping.}
    \end{table*}
    
    \subsection{Limiting magnitude in our images}
    \label{limmag}
    To estimate the noise level on the final best Sgr~A* removed residual image per month, we select the central 74~mas $\times$ 74~mas region on each image, which corresponds to the FWHM of the fiber FOV and also blank out the 5 $\times$ 5 mas region around the peak of our newly detected star. We then quote the root-mean-square (rms) calculated in this region as the 1-$\sigma$ noise level for the residual image. S2 is located outside of this region, so it does not affect the noise calculation here. We report these numbers in Table~\ref{t2}. As the flux in our images is normalized to S2, so is the 1-$\sigma$ noise level. We find a mean noise level of $1.7 \pm 0.2 \times 10^{-3}$ in relative brightness, corresponding to $m_\mathrm{K} = 21.0 \pm 0.2$. This is the statistical limit corresponding to the brightness of a star that would create a peak in our images comparable to the noise floor in the absence of systematics. For a 5-$\sigma$ detection, an object would need to be at least as bright as $m_\mathrm{K} = 19.3 \pm 0.2$ if no systematics were present. We note here that these numbers are "apparent" magnitudes in the sense that no fiber damping effects have been considered. Therefore, the real magnitude of the star for a 5-$\sigma$ detection will be between 18.5 and 19.3 magnitude, depending on the distance to the center of the field. 

    \subsection{Interferometric model fit of the data}
    
    Since the star we detected is just above the 5-$\sigma$ detection level, we use the Meudon model-fitting code as an independent way of verifying that our detection is real. This code was already used in our papers on the orbit of S2 \citep{2018A&A...615L..15G} and is able to fit two or three sources.
    
    Our image reconstruction relies on the complex visibilities, i.e. amplitude and phase. The model fit, on the other hand, uses closure amplitudes and phases, giving equal weight to each.

    We select the same frames as for the image reconstruction and fit the frames from each night together, but separate the two polarisation states. 
    For each night, we fit one flux per file for Sgr~A*, one flux per file and
    per telescope for S2, a single background flux, a spectral index for Sgr~A*, and a single position for S2 and the third source respectively
    (neglecting proper motion over the course of the night). S2, the third source, and the background are assumed to have a common spectral index, which is fixed.

    Blindly starting such a fit would very likely not find the best minimum, and hence we fed the fitting routine with starting values derived from our imaging. We applied a simple $\chi^2$ minimization algorithm to explore the parameter space in the $\pm$ 10~mas region around the position of the star derived from imaging. 
    
    In all cases, the fits recovered the third star. The reduced $\chi^2$ is of the order 2-3. Comparing the Bayesian information criterion (BIC) of the fits with and without the third source strongly supports its presence, and the third source is consistently found at the position revealed by the imaging reconstruction to within the uncertainties. For example, for the night of April 18th, 2019, using polarization state P1, the triple source model yields a reduced $\chi^2$ of 2.1 and a BIC of 493 while the binary model yields a reduced $\chi^2$ of 2.9 and a BIC of 664. $\Delta$BIC is thus 171. The significance of the detection estimated as flux over uncertainty of the third source is 9.6 $\sigma$. The polarization P2 yields similar statistical estimators. We note the improvement by fitting three stars instead of two stars is not obvious within a 5-minute frame but only for a full night of data.

    \subsection{Further checks}
    
    To further verify our detection, we also reconstruct images for each (linear) polarization channel of GRAVITY separately. Our newly detected star appears in both channels, indicating an origin from the sky rather than a noise artifact.
    
    We further run the CLEAN process on Sgr~A* (for which the spectral index is only poorly constrained) using a spectral index of zero, and we can still recover the newly detected star in the residual images. Since we only remove Sgr~A*, any difference between using different spectral indices must come from sidelobes and beam patterns related to Sgr~A*. This is especially helpful when there are only fewer than ten frames available for a certain night. 
    
    We can also rule out that the detection is caused by the overlap of sidelobes from S2 and Sgr~A*. As shown in the appendix, three out of six baseline patterns are oriented in the southeast-northwest direction and one baseline pattern is oriented in the northeast-southwest direction. In principle, these baseline patterns could mimic a detection. However, since S2 is moving away from Sgr~A*, any fake target formed in this way should also move away from Sgr~A* in parallel to S2's motion. This is inconsistent with the star showing up at (almost) the same position over five months.

    
    \subsection{Motion on the sky of the star}
    
\begin{figure}
   \centering
   \includegraphics[width=\hsize]{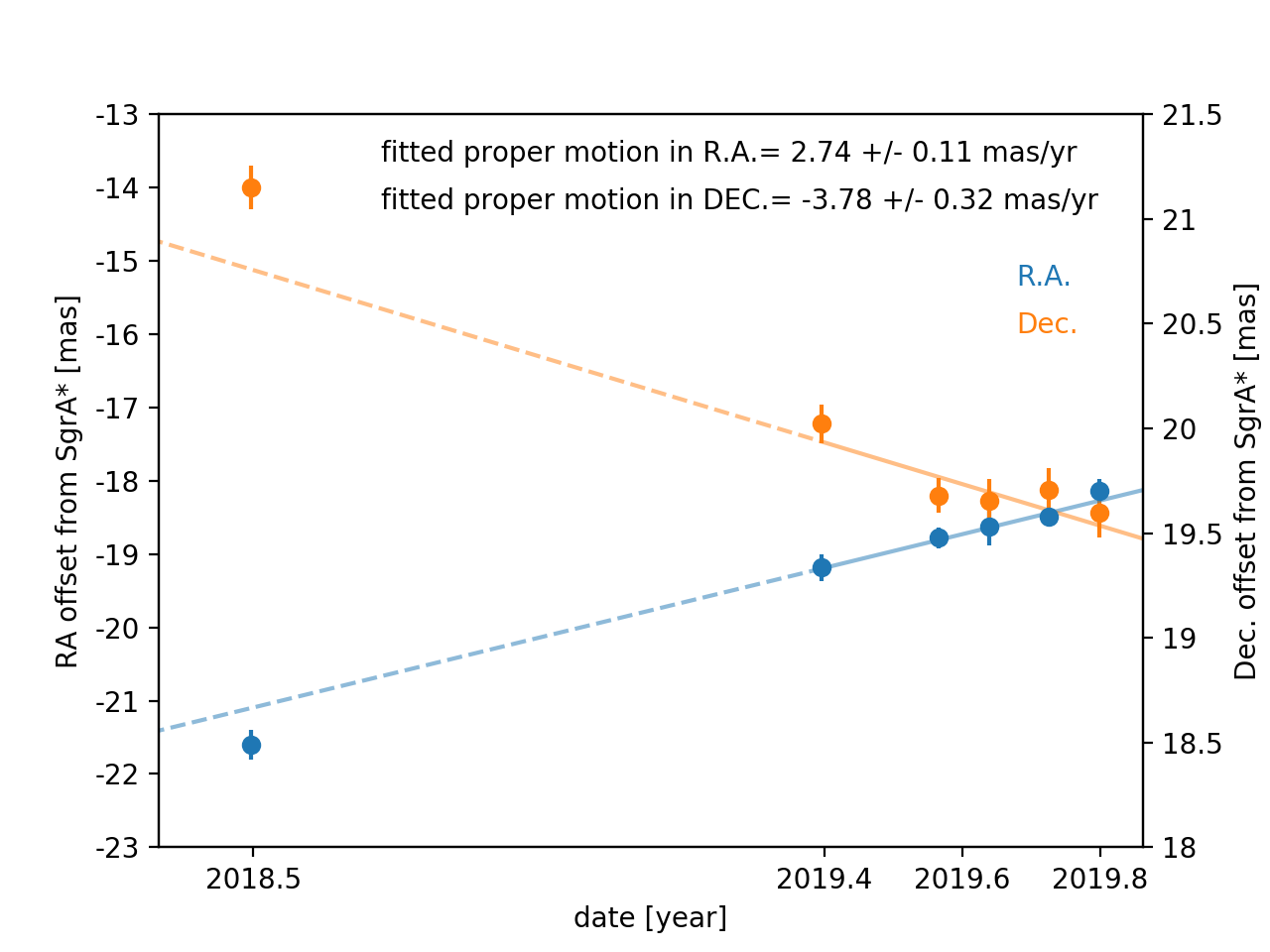}
      \caption{Fitted positions of our tentative detected star relative to Sgr~A* with time. The fitting was done only with the 2019 data points, as shown by the solid part of the lines, and then extrapolated to the 2018 epoch, as indicated by the dashed line. 
              }
         \label{figure:proper-motion}
   \end{figure}

    We show the change of position of our newly detected star as a function of time in Fig.~\ref{figure:proper-motion}. We see regular changes in both R.A.\ and Dec. with time, and a simple linear fitting (shown in solid line) gives a proper motion in R.A.\ of $2.38 \pm 0.29\,$mas yr$^{-1}$ and in Dec.\ of $-2.74 \pm 0.97\,$mas yr$^{-1}$. 
    
    We list the proper motions measured based on the positions derived from the model fitting method in Table~\ref{t3}. These results are within 4-$\sigma$ uncertainties of the image-based results. 
    
    \begin{table}
    \caption{\label{t3}Fitted proper motion for the detected star}
    \centering
    \begin{tabular}{lcccc}
    \hline\hline
    Method/data & p$_{R.A.}$  &$\delta$p$_{R.A.}$  &p$_{Dec}$ &$\delta$p$_{Dec}$  \\
    \hline
    imaging (2019)      & 2.38      &0.29   &  -2.74    & 0.97 \\
    model fitting P1    & 3.36      &0.82   &  -4.44    & 1.33 \\
    model fitting P2    & 3.33      &0.49   &  -4.08    & 1.17 \\
    \hline
    imaging 2019 + 2018 & 2.74 & 0.11 & -3.78 & 0.32 \\
    \hline
    \end{tabular}
    \tablefoot{Fitted proper motion of the detected star, from both imaging and model fitting results. Here P1 and P2 stand for the two different polarization from the GRAVITY data. All the units for proper motion and uncertainties are in mas $\mathrm{yr}^{-1}$. The bottom row includes also the position of the tentative detection from the 2018 data (sec.~\ref{tent2018}).}
    \end{table}
    
    \subsection{Tentative detection from 2018 data}
    \label{tent2018}
\begin{figure}
   \centering
   \includegraphics[width=0.5\hsize]{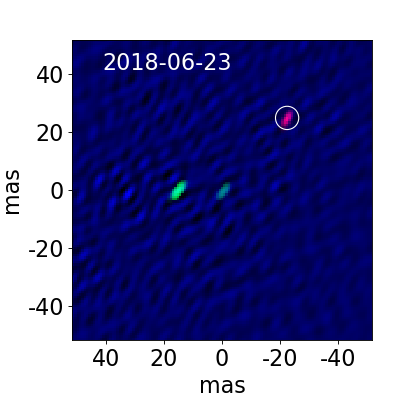}
      \caption{Three-color composite image for the central 100 mas region around Sgr~A* for the night 2018-06-23. Sgr~A* (located at the center of the image) and S2 are color-coded in green, our tentative detection is shown in red (marked with a white circle), and the residual map is shown in blue. North is up and east to the left.
              }
         \label{figure:S62-2018}
   \end{figure}

    Due to the relatively small proper motion of our newly detected star, we predict that it is also located inside our field of view in the 2017 and 2018 data. We have re-checked the images reconstructed from the 2017 and 2018 data \citep{2018A&A...615L..15G} and noticed a tentative detection near the extrapolated position at the $\sim$ 3$\sigma$ level in one epoch (2018.48, 2018-06-23). We show the reconstructed three-color image in Fig.~\ref{figure:S62-2018}. The measured position offset of this tentative detection to Sgr~A* is -21.6 $\pm$0.2 mas in R.A.\ and 24.0 $\pm$0.3 mas in Dec. We also refit the proper motion with both 2018 and 2019 positions and find a value of $2.74 \pm 0.11\,$mas yr$^{-1}$ in R.A.\ and  $-3.78 \pm 0.32\,$mas yr$^{-1}$ in Dec. 
    
\section{Discussion}

    In this section we discuss the nature of the object that has been detected in section 5.1 and 5.2, and we illustrate our current limitation of the imaging technique in section 5.3.

    \subsection{Identification of the detected object and constraints on the 3D position}
    
    With the position on the sky and the proper motion derived for the detected star, we cross-check whether any of the known S-stars in the Galactic Center region can be identified with this object. Inspecting the search map in figure~\ref{figure:naco-mosaic} shows that only a handful of stars could potentially be the detected star. Combining the adaptive-optics based astrometry with the new positions leads to a clear best match: S62. We show the on-sky position of this star from previous NACO images and from our results together in Fig.~\ref{figure:S62-orbit}. 
    
    Since S62 is at least in projection close to Sgr~A*, it is worth asking why the motion appears to be linear with constant velocity. Given the projected separation of $\approx 27\,$mas and the lack of observed acceleration, we can derive a lower limit on the $z$ coordinate along the line of sight. At a distance of $|z| < 195\,$mas, we would have detected an acceleration with $>3\sigma$ significance, and at $|z| < 161\,$mas the significance would have reached $>5\sigma$. We conclude that S62 resides at $|z| > 150\,$mas. Further data will of course either improve this limit or eventually detect an acceleration and thus determine $|z|$.
    
    \subsection{S62 and S29}
    
            \begin{figure*}
     \centering
     \includegraphics[width=\hsize]{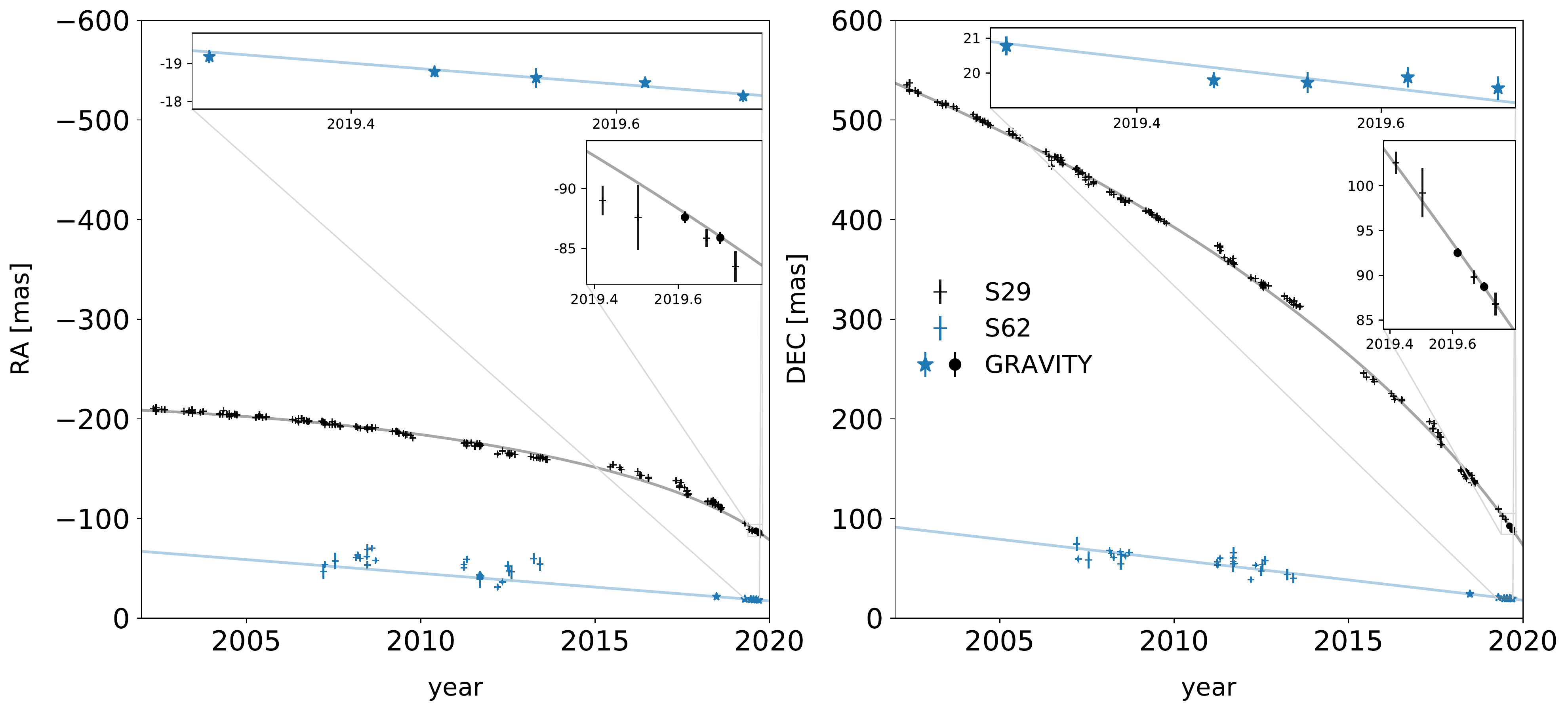}
      \caption{Change of on-sky position for both S62 and S29 in R.A. (left panel) and Dec. (right panel) vs. time as observed by GRAVITY and from previous VLT NACO measurements. In each panel, measurements for S62 are shown in blue and measurements for S29 are shown in black. New measurements from GRAVITY are shown in stars (imaging) and filled circles (dual-beam astrometry), while previous NACO measurements are shown in crosses. The solid lines for S29 show the best-fit orbit; While the S62 data is insufficient to constrain any orbital parameters yet, we show a linear fit to the GRAVITY data. The two insets show the zoom-in view of the GRAVITY data points in 2019 for S62 and S29, respectively. 
              }
         \label{figure:S62-orbit}
   \end{figure*}
   
            \begin{figure}
     \centering
     \includegraphics[width=\hsize]{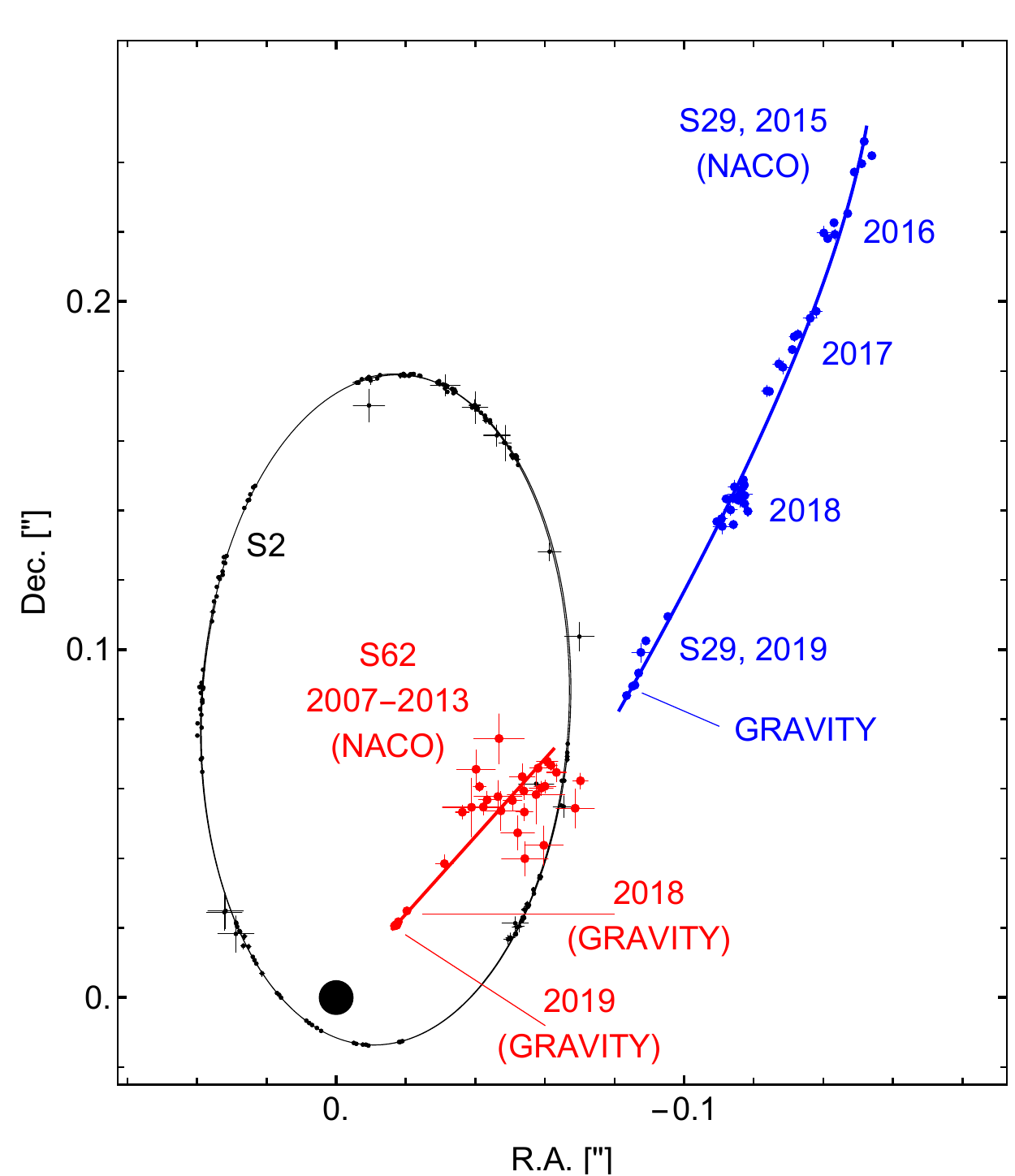}
      \caption{The fitted orbit of S2(black), S62(red) and S29(blue). Sgr~A* is located at [0,0] indicated by a black filled circle. 
              }
         \label{figure:noise}
   \end{figure}

    Our data and identification are inconsistent with the conclusions of \cite{Peissker}, who reported previously an orbit for S62 with a 9.9-year period and extreme eccentricity. Their proposed orbit predicts a position of (RA, Dec)$\,=\,(\mbox{-89} .. \mbox{-85}, \mbox{+98} .. \mbox{+95})\,$mas over the time span of our observations, with a proper motion of $\approx (+7,-5)\,$mas~$\mathrm{yr}^{-1}$. The position is outside the FOV of our GRAVITY data centered on Sgr~A* and incompatible with our detection reported here. Hence, we address the question what object \cite{Peissker} have actually seen. 
    
    To check the position proposed by \cite{Peissker}, we pointed GRAVITY to the position in question on 13, 14 August 2019 and 13 September 2019. In all three epochs, we find a single dominant source in the field of view, for which we can derive dual-beam positions (for the methodology see \citealt{2020A&A...636L...5G}). We verify that what we observed is a celestial source by offsetting our science fiber pointing position by 10 mas to the east and north separately on August 14th, and we do see the single dominant source appearing west and south of the image center by 10~mas. The observed target is about 10 times fainter than S2, so $m_{K} \approx$ 16.6 mag. The August position is $(-87.6, 92.5)\,$mas with an uncertainty around 0.1~mas. In September, the object was at $(-85.9, 88.7)\,$mas with a similar uncertainty. The resulting proper motion is thus $\approx (+21,-46)\,$mas~$\mathrm{yr}^{-1}$, where the errors are in the $1\,$mas~$\mathrm{yr}^{-1}$ regime. This object moves thus much faster than what \cite{Peissker} predict on basis of their proposed orbit for the object they assume was S62 near its apocenter. 
    
    The positions and proper motions from GRAVITY match perfectly the orbital trace of the star S29 \citep{2017ApJ...837...30G}, as shown in Fig.~\ref{figure:S62-orbit}. We conclude that the offset pointing in August and September 2019 with GRAVITY has seen S29, but we cannot report the detection of any object that would correspond to the 9.9-year orbit claimed by \cite{Peissker}.

    With a much higher spatial resolution from our data, we are not able to confirm the findings of \cite{Peissker}, and we can only speculate whether possibly misidentifying S29 with S62 in the lower resolution NACO images of the past few years yields the 9.9-year orbit.

    \subsection{Limitations of our imaging technique}
    \label{limittech}
    In section~\ref{limmag} we showed that our current detection limit is at the $\approx$ 5$\sigma$ level on each image. To see if and how we can push our detection limit deeper, we discuss here the current limitations in our imaging technique. 
    
    We illustrate this in Fig.~\ref{figure:noise}. First, we compare the dynamic range reached on the CLEANed image between S2 and Sgr~A* frames in panel a. Here we calculate the dynamic range as the ratio between the peak flux in each CLEANed image and the rms noise level in the central 74~mas $\times$ 74~mas region on each residual image. Each frame is 320 seconds long. Since there are no known bright stars within the FOV of S2 and we also do not see Sgr~A* in these frames, we can use these S2 frames to show the dynamic range that can be reached with the CLEAN method. We use the same parameter settings in CLEAN for S2 as for Sgr~A* frames, except we set the spectral index to zero for S2. We don't reach the same dynamic range level as for S2 frames. We then compare the noise level reached between S2 frames and Sgr~A* frames in panel b, in which Sgr~A* frames are slightly deeper. This suggests the "noise cap" we reach is from the data rather than the CLEAN algorithm itself. Since we know that both S2 and S62 are present in the Sgr~A* removed frames, we further check if they could be the limiting factor for the dynamic range for Sgr~A*. We show the distribution of the peak flux in all the Sgr~A*-removed residual frames in panel c for comparison. Here the brightness is normalized to S2. As listed in Table~\ref{t2}, the relative brightness for S62 is around 0.01, which is below most of the residual peaks we see here. We have checked the position of these residual peaks and almost none of them are centered on Sgr~A* or the adjacent sidelobes, so we can say the residual peaks are not a result of Sgr~A* not being completely removed. Although combining multiple residual images over a night can already smooth out some of these residual peaks, we explore the source of these peaks below.
    

         \begin{figure*}
     \centering
     \includegraphics[width=\hsize]{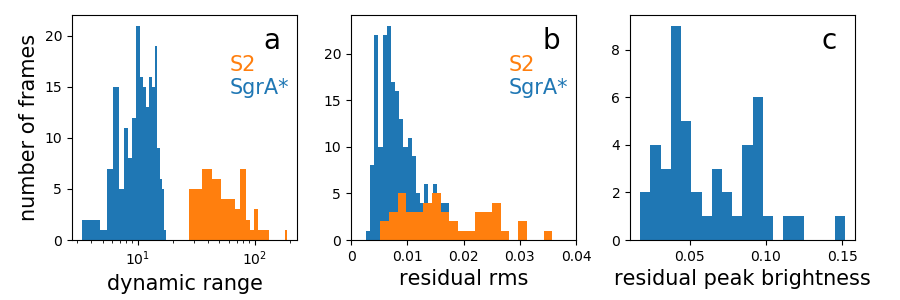}
      \caption{The noise behavior of our data. (a) A comparison of the dynamic range we can reach between S2 and Sgr~A* frames. The dynamic range is calculated as the ratio between the peak flux in each CLEANed image versus the rms in the central 74~mas $\times$ 74~mas region on the residual map. (b) The distribution of the rms noise in all the Sgr~A*-removed residual frames and S2 frames. (c) The distribution of the peak flux in all the Sgr~A* removed residual frames. For comparison, S62 has a relative brightness of 0.01. These numbers are normalized to S2 flux. 
              }
         \label{figure:noise}
   \end{figure*}
   
   
      \begin{figure}
   \centering
   \includegraphics[width=10cm]{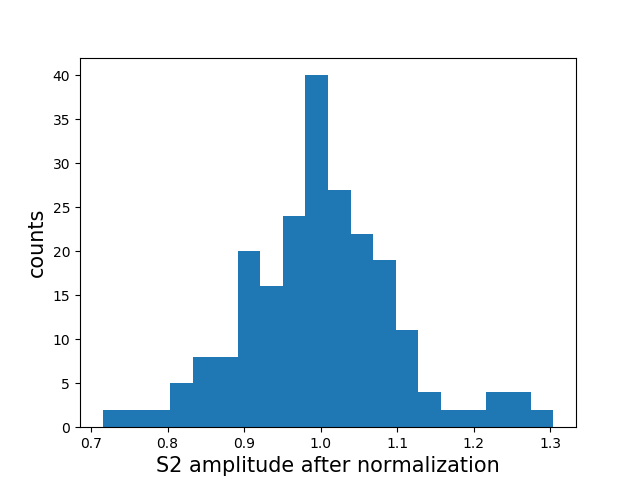}
      \caption{The flux distribution of S2 frames, which shows the uncertainty of our flux calibration.
              }
         \label{figure:s2-amp-unc}
   \end{figure}
    

    Due to the nature of the CLEAN method, these high residual peaks indicate a mismatch between the dirty image and the dirty beam used in the deconvolution. Following the traditional radio interferometry definition, the dirty beam is simply the Fourier transformation of the observation u-v coverage, which does not account for any distortion arising from errors in the real measurements. In other words, the mismatch could be put as calibration residuals of the measured visibility. As mentioned above in our data reduction process, the calibration is done by normalizing the Sgr~A* complex visibility with that from an S2 frame. The visibility phase zero point is the position of S2 plus the pre-entered separation between Sgr~A* and S2, which are calculated based on the S2 orbit. Any phase error here would affect all six baselines together. Meanwhile, our visibility amplitude has accounted for the photometric flux of each baseline from each telescope pair, which is a telescope-dependent time-variable quantity. We suspect that any variability here would introduce inconsistency of the visibility amplitude between different baselines, which cause deviations from our calculated synthesis beam. We show how well the amplitude of S2 is calibrated in Fig.~\ref{figure:s2-amp-unc}. Ideally, the amplitudes should all be exactly unity; in practice the majority of our data has a deviation of the amplitude within 10$\%$.     
    
    Another known systematic effect in our instrument is the field-dependent phase error, as reported in \citep{2020A&A...636L...5G}. This error will introduce a position offset of any detected targets away from the field center (e.g. at the position of S62, for a typical 40 $\deg$ phase error, the position offset is between 0.2 and 0.4 mas). Additionally, this error will reduce the brightness of any targets, as different baselines will shift the target in different directions, thus decreasing the total coherence. Equivalently this effect distorts the measured PSF shape from the theoretical one. Although here we only clean on Sgr~A*, which is at the field center and not affected, this effect needs to be taken into account in the image reconstruction stage, prior to CLEAN. In the analysis chain presented here, there is no way to include such a systematic effect, since it needs to be taken into account at the level of the van-Cittert-Zernike theorem \citep{TMS}. In essence, this means one would need to write a dedicated imaging code, which is beyond the scope of this work.
    
     The low noise levels we have achieved show that further analysis in this direction might be very fruitful and can potentially reveal stars at magnitudes fainter than S62. Our hope is that a few of these reside physically close to Sgr~A*, such that they can be used as relativistic probes of the spacetime around the MBH.


\section{Conclusions}
    In this paper we report the detection of a $m_{K}=$ 18.9 magnitude star within 30 mas projected distance to Sgr~A* from recent GRAVITY observation of the Galactic Center region. We have successfully adopted the CLEAN algorithm used in radio interferometry to near-infrared interferometry. Our method recovers the faint star in multiple epochs across different months. This detection is further confirmed by a model-fitting method which uses squared visibilities and closure phases. We also measure the proper motion of this faint star to be 2.38 $\pm$ 0.29 mas $\mathrm{yr}^{-1}$ in R.A.\ and -2.74 $\pm$ 0.97 mas $\mathrm{yr}^{-1}$ in Dec. By comparing the orbits of previously known S-stars, we identify our source with the star S62 as reported in \cite{2017ApJ...837...30G}. Throughout our observation we also detect S29 within 130 mas of Sgr~A*. We are not able to confidently identify other faint stars in our current images, probably due to the limitation of our calibrations. We expect future upgrades of the instrument in the framework of the GRAVITY+ project and better calibration and removal of the PSF from bright stars will lead to deeper images near Sgr~A*.



\begin{acknowledgements}
      We thank the referee for a helpful report. F. G. thanks W. D. Cotton and E. W. Greisen for the help on AIPS and UVFITS data format. We are very grateful to our funding agencies (MPG, ERC,
CNRS [PNCG, PNGRAM], DFG, BMBF, Paris Observatory [CS, PhyFOG],
Observatoire des Sciences de l’Univers de Grenoble, and the Fundação para
a Ciência e Tecnologia), to ESO and the ESO/Paranal staff, and to the many
scientific and technical staff members in our institutions, who helped to
make NACO, SINFONI, and GRAVITY a reality. S. G. acknowledges the
support from ERC starting grant No. 306311. F. E. and O. P. acknowledge the
support from ERC synergy grant No. 610058. A. A., P. G., and V. G. were
supported by Fundação para a Ciência e a Tecnologia, with grants reference
UIDB/00099/2020 and SFRH/BSAB/142940/2018.
\end{acknowledgements}

\appendix
\section{interferometric beam}
\begin{figure*}
   \centering
   \includegraphics[width=10cm]{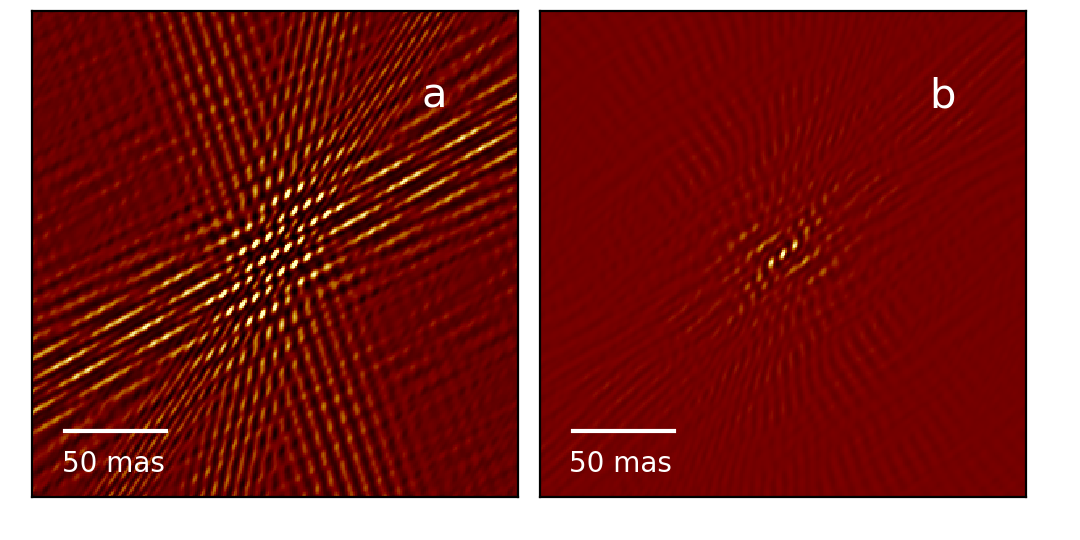}
      \caption{(a) GRAVITY interferometric beam shape after 5-minutes exposure. (b) interferometric beam of combining whole nights' exposure. 
              }
         \label{figure:append-beam}
   \end{figure*}

\begin{figure*}
   \centering
   \includegraphics[width=10cm]{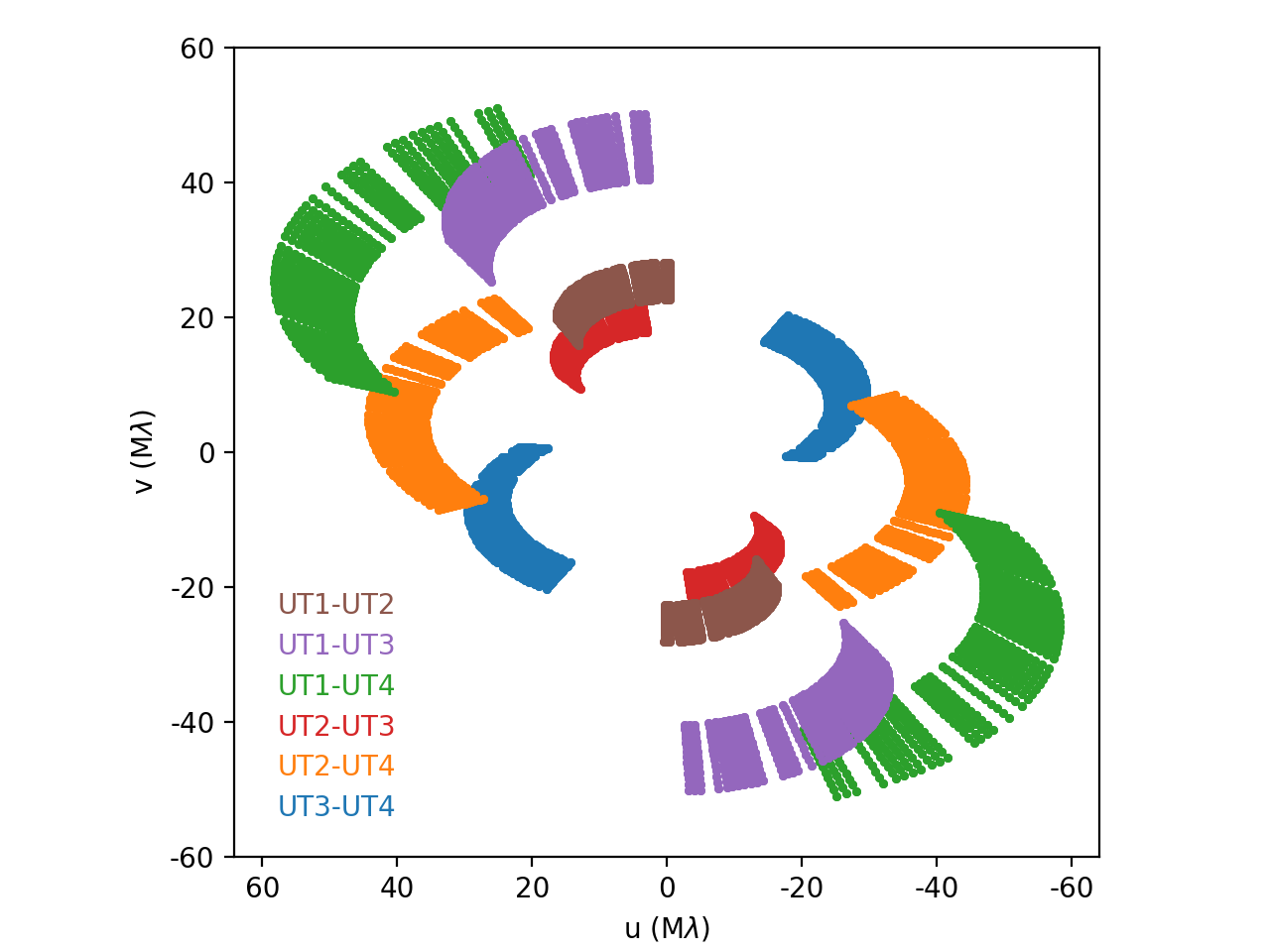}
      \caption{The uv-coverage of the VLTI UT telescopes from the night of 2019-07-17. 
              }
         \label{figure:uvcoverage}
   \end{figure*}

We show in Fig.~\ref{figure:append-beam} both the GRAVITY interferometric beam over a 320-second integration and over a whole night. The uv coverage of a typical night is shown in Fig.~\ref{figure:uvcoverage}.

%
%

\end{document}